\documentclass[amsmath,twocolumn,showpacs,aps,prl]{revtex4}
\usepackage{bm}
\usepackage{graphicx}
\usepackage{epsfig}

\begin{document}

\begin{abstract}
The interplay of electron-electron interactions and spin-orbit
coupling leads to a new contribution to the homogeneous optical
conductivity of the electron liquid. The latter is known to be
insensitive to many-body effects for a conventional electron system
with parabolic dispersion. The parabolic spectrum has its origin in
the Galilean invariance which is broken by spin-orbit coupling. This
opens up a possibility for the optical conductivity to probe
electron-electron interactions. We analyze the interplay of
interactions and spin-orbit coupling and obtain optical conductivity
beyond RPA.
\end{abstract}


\title{Optical Conductivity of a 2D Electron Liquid with Spin-Orbit Interaction}

\author{Abdel-Khalek Farid}
\affiliation{Department of Physics, University of Utah, Salt Lake
City, UT 84112}

\author{Eugene G. Mishchenko}
\affiliation{Department of Physics, University of Utah, Salt Lake
City, UT 84112} \maketitle

{\it Introduction}. Spin-polarized transport phenomena have recently
become a subject of extreme interest, with the ultimate goal of
achieving selective local manipulation of spins by means of electric
fields.  The vast majority of theoretical works and spintronic
proposals, however, utilizes the approximation of independent
electrons, neglecting many-body effects. Effects of interactions,
however, are traditionally among the most interesting, though most
challenging, problems in condensed matter physics.
 In this Letter we report the effect that arises as a result of the {\it
 interplay}
 of electron-electron interactions and spin-orbit coupling in an electron liquid.

The response of an electronic system to a homogeneous electric field
is described by its optical conductivity $\sigma(\omega)$. This
quantity is known to be {\it independent} of the effects of
electron-electron interactions for a system with the parabolic
dispersion, $H=p^2/2m$, as long as collisions with impurities,
surface imperfections and phonons can be neglected \cite{Pines&Noz}.
This is due to the fact that electric current, ${\bf j}=e\sum {\bf
p}/m$, is proportional to the total momentum of particles. The
latter, however, is not changed by electron collisions in a
translationally invariant system (that implies absence of Umklapp
scattering, usually negligible in semiconductors), which also
includes the presence of a homogeneous electric field. Therefore,
homogeneous optical conductivity typically {\it cannot} be used as a
probe for many-body effects. The situation changes completely in the
presence of spin-orbit coupling.

The parabolicity of the spectrum is intimately related to the
Galilean invariance. However, in semiconductors such as GaAs or
InAs, spin-orbit coupling is always present, being especially
pronounced in two-dimensional structures transversally confined to
quantum wells. Spin-orbit coupling is  relativistic in nature and
breaks Galilean invariance, making many-body effects important for
the optical conductivity $\sigma(\omega)$. Indeed, in the presence
of spin-orbit coupling in the Hamiltonian, $ H_{so} =p^2/2m-{\bf
h}_{\bf p} \cdot \hat{\bm \sigma}$, the operator of electric
current, ${\bf j}=e\sum [{\bf p}/m -\nabla_{\bf p} ( {\bf h}_{\bf p}
\cdot \hat{\bm \sigma})]$, becomes spin-dependent and does not
reduce to the total momentum. The conservation of the latter during
electron-electron scattering events no longer implies conservation
of current. This makes the homogeneous optical conductivity {\it
sensitive to many-body effects}.

 Though our method
is applicable for arbitrary spin-orbit interaction, we concentrate
here on its isotropic, ``Rashba'', type \cite{BR}, which assumes
${\bf h}_{\bf p} = \lambda(-p_y, p_x,0)$. The effective
momentum-dependent magnetic field ${\bf h_p}$ lies within the plane
of 2DEG while being perpendicular to the electron momentum. Lifting
of the spin degeneracy due to spin-orbit coupling results in the
possibility of single-particle absorption (Landau damping) even for
zero transferred momentum $q$. This leads to the box-like shape
contribution into the real part of the optical conductivity at zero
temperature \cite{MCE,MH} (hereinafter we assume $\hbar=1$),
\begin{equation}
\label{RPA} \sigma'_1(\omega)=\frac{e^2}{16}~ \Theta
(2m\lambda^2-|\delta \omega|), ~~~\delta \omega =\omega-2\lambda
p_F,
\end{equation}
where $p_F$ is the value of the Fermi momentum. Spin-orbit induced
Landau damping (\ref{RPA}) is also known as the ``combined''
\cite{RSh} or ``chiral spin'' \cite{ShKhF} resonance. The issue of a
modification of the chiral spin resonance by electron-electron
interactions has been addressed with the help of the Landau
interaction function formalism \cite{ShKhF}. Though within this
model interactions renormalize the effective strength of the
spin-orbit coupling constant (see also the earlier paper
\cite{ChR}), they do not result in the broadening of the chiral spin
resonance.

It is the aim of our work to analyze the many-body effects {\it
beyond} random phase approximation, Hartree-Fock model or Landau
interaction function formalism. In particular, we are interested in
the absorption channel that involves the excitation of two
electron-hole pairs. Taking into account two-pair processes removes
the phase-space constraint that leads to the $\Theta$-function in
the single-pair term, Eq.~(\ref{RPA}), and, thus, results in a much
broader contribution. Indeed, constraints in the single-pair channel
originate from the vanishing of the total transferred momentum $q$
in case of a homogeneous external electric field. In contrast,
two-pair processes have a large phase space available, since two
pairs can carry large momenta of opposite signs, and still have zero
net momentum. To calculate the contribution from the two-particle
channel to the optical conductivity, one needs to evaluate the
non-RPA diagrams shown in Fig.~1. For a finite temperature and the
simplest case of a short-range interaction independent of momentum,
$V$, we obtain,
\begin{equation}
\label{result} \sigma'_2(\omega)=\frac{2e^2m^2\lambda^2
V^2}{3v_F^2(2\pi)^4 \omega^2}
\left\{ \begin{array}{cc} 2\omega^2, & \omega \gg \pi T,\\
(2\pi T)^2, & \omega \ll \pi T,
\end{array} \right.
\end{equation}
where $v_F=p_F/m$ is the Fermi velocity. The subscript in the
notation of $\sigma'_2(\omega)$ in Eq.~(\ref{result}) distinguishes
the many-body contribution from the single-pair result,
Eq.~(\ref{RPA}).
\begin{figure}
\includegraphics[width=0.35\textwidth]{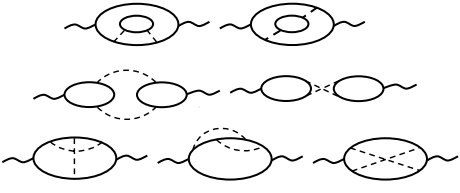}
\caption{Non-RPA contributions into optical conductivity from the
two-pair channel. Dashed line stands for the electron-electron
interaction. The last three diagrams originate from exchange
processes.}
\end{figure}

{\it Derivation}. Instead of calculating the real part of the
optical conductivity directly from the set of diagrams of Fig.~1, we
employ an equivalent and arguably more transparent method. Namely,
we identify various processes leading to absorption (and emission)
in the two-particle channel and calculate their transition
probabilities using the Golden rule formalism \cite{Mishch}. In
diagrammatic terms, these various processes correspond to all
possible cuts of the diagrams, Fig.~1, across any four fermion
lines.

We begin with calculating transition rates between different
two-particle states. The two-particle wave function is given by the
Slater determinant,
\begin{equation}
\psi^{ab}_{\bf pk}({\bf x}_1,{\bf
x}_2)=\frac{1}{\sqrt{2}}\left[\psi^a_{\bf p}({\bf x}_1)\psi^b_{\bf
k}({\bf x}_2)-\psi^a_{\bf p}({\bf x}_2)\psi^b_{\bf k}({\bf x}_1)
\right].
\end{equation}
Here $\psi^a_{\bf p} ({\bf x})$ is a single-particle wave function
with the momentum ${\bf p}$ belonging to the $a$-th spin subband, $a
=\pm 1$; the notation ${\bf x}$ stands for the in-plane coordinates,
${\bf x}=(x,y)$. For the ``Rashba'' coupling the eigenstates are
\begin{equation} \label{states} \psi^a_{\bf p} ({\bf x})
=\frac{1}{\sqrt{2}} \left(
\begin{array}{c} e^{i\chi_{\bf p}/2}
\\ a e^{-i\chi_{\bf p}/2}
\end{array} \right) e^{i{\bf p \cdot x}/\hbar},
\end{equation}
where $\chi_{\bf p}$ denotes the angle between the momentum ${\bf
p}$ and the $y$-axis. The energy of these eigenstates is
\begin{equation}
\label{spectrum} \epsilon^a_{{\bf p}} = {p^2}/{2m}+a\lambda p.
\end{equation}
It is now necessary to calculate the probability of a transition
from a state $\psi^{ab}_{\bf pk}$ into another state $\psi^{cd}_{\bf
p'k'}$ in the presence of both the electron-electron interaction
$V({\bf x_1}-{\bf x_2})$ and the electric field, which is described
by a scalar potential,
\begin{equation}
\label{phi} \phi ({\bf x},t) = \phi_0 e^{-i\omega t +i{\bf q \cdot
x}}+ \phi_0^* e^{i\omega t -i{\bf q \cdot x}}.
\end{equation}
Coupling of electrons to the external field (\ref{phi}) as well as
the electron-electron interaction are treated in the second-order
perturbation theory. Spin-orbit coupling, on the other hand, {\it is
not assumed} to be small for the time being. Transition probability
between different two-particle states, accompanied by the absorption
of the energy $\omega$ from the external field, has the following
form,
\begin{eqnarray}
\label{transition} dW^{ab\to cd}_{{\bf pk} \to {\bf p'k'}}
 = 2\pi e^2|\phi_0|^2  \delta (\epsilon^a_{\bf
p}+\epsilon^b_{\bf k}-\epsilon^c_{\bf p'}-\epsilon^d_{\bf k'} + \omega)\nonumber\\
\times \vert \sum_f {\cal M}_f\vert^2 \delta({\bf p}+{\bf k}-{\bf
p'}-{\bf k'} + {\bf q}) \frac{d^2p' d^2k'}{(2\pi )^2}.
\end{eqnarray}
where  ${\cal M}_f$ is the amplitude  for the transitions that occur
via virtual states belonging to a subband $f$,
\begin{widetext}
\begin{eqnarray}
\label{matrix}
{\cal M}_f=\frac{{\cal A}^{af}_{\bf p,p+ q} \Bigl({\cal A}^{fc}_{\bf
p+ q,p'} V_{\bf k-k'}{\cal A}^{bd}_{\bf k,k'}-{\cal A}^{fd}_{\bf p+
q, k'}V_{\bf p'-k}{\cal A}^{bc}_{\bf k,p'}\Bigr)}{\epsilon^a_{\bf p}
-\epsilon^f_{\bf p+ q}
 + \omega }
+\frac{\Bigl({\cal A}^{af}_{\bf p,p'- q} V_{\bf k-k'}{\cal
A}^{bd}_{\bf k,k'}-{\cal A}^{ad}_{\bf p, k'}V_{\bf p-k'}{\cal
A}^{bf}_{\bf k,p'-q}\Bigr) {\cal A}^{fc}_{\bf p'-
q,p'}}{\epsilon^c_{\bf p'}-\epsilon^f_{\bf p'-
q}-\omega} \nonumber\\
+\frac{{\cal A}^{bf}_{\bf k,k+ q} \Bigl({\cal A}^{ac}_{\bf p,p'}
V_{\bf p'-p}{\cal A}^{fd}_{\bf k+ q,k'}-{\cal A}^{ad}_{\bf p,
k'}V_{\bf p-k'}{\cal A}^{fc}_{\bf k+ q,p'}\Bigr) }{\epsilon^b_{\bf
k}-\epsilon^f_{\bf k + q} + \omega} +\frac{\Bigl({\cal A}^{ac}_{\bf
p,p'} V_{\bf p-p'}{\cal A}^{bf}_{\bf k,k'-q}-{\cal A}^{af}_{\bf p,
k'- q}V_{\bf p'-k}{\cal A}^{bc}_{\bf k,p'}\Bigr) {\cal A}^{fd}_{\bf
k'- q,k'} }{\epsilon^d_{\bf k'}-\epsilon^f_{\bf k'- q}- \omega }.
\end{eqnarray}
\end{widetext}
Here $V_{\bf p-p'}$ stands for the Fourier transform of the
interaction potential, the notation ${\cal A}^{ac}_{\bf p,p'}$ is
used for the overlap of spin wave functions of single-electron
states (\ref{states}) before ($\psi^a_{\bf p}$) and after
($\psi^c_{\bf p'}$) the scattering:
\begin{equation}
\label{projector}
 {\cal A}^{ac}_{\bf p,p'}=\frac{1}{2}
\left(e^{i(\chi_{\bf p}-\chi_{\bf p'})/2}+ac~ e^{-i(\chi_{\bf
p}-\chi_{\bf p'})/2}\right).
\end{equation}
The origin of various terms in the transition probability
(\ref{transition})-(\ref{matrix}) is graphically represented in Fig.~2a).

The knowledge of transition probabilities allows one to find the
rate at which the electron system absorbs energy from the external
field (\ref{phi}). Taking into account the population of the
electronic states, the energy absorption rate can be written as the
sum over initial and final states,
\begin{equation}
\label{dissip} I_{\rm abs}=  \frac{\omega}{4} \sum_{abcd} \int
\frac{d^2p d^2k}{(2\pi)^4}~ dW^{ab\to cd}_{{\bf pk} \to {\bf p'k'}}
n^a_{p}n^b_{ k}(1-n^c_{p'})(1-n^d_{k'}). \nonumber\\
\end{equation}
Here $n^a_p$ is the Fermi-Dirac distribution for the $a$-th subband;
the coefficient ${1}/{4}$ prevents double-counting of the initial
and final states. The rate of emission of energy, $I_{\rm em}$ is
most simply found from the detailed balance principle \cite{LLVIII},
$I_{\rm em}=I_{\rm abs}e^{-\omega/T}$. The energy dissipation rate
$I_{\rm abs}-I_{\rm em}$ can then be related to the real part of the
optical conductivity at finite $q$,
\begin{equation}
\label{dissipphon1} \sigma'(\omega,q)=\frac{(1-e^{-\omega/T})I_{\rm
abs}}{2q^2|\phi_0|^2}.
\end{equation}
\begin{figure}
\includegraphics[width=0.4\textwidth]{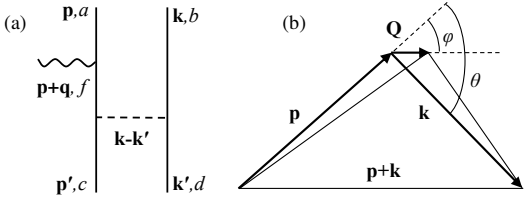}
\caption{a) Graphic representation of the two-electron absorption
amplitude ${\cal M}_f$, see Eq.~(\ref{matrix}). The wavy line
(external electric field) can be inserted in four different ways
(not shown here). In addition, interchange of the final states
$c{\bf p'} \leftrightarrow d{\bf k'}$ yields four exchange terms,
leading to the total of eight different terms in Eq.~(\ref{matrix}).
Propagators of the virtual states result in the energy denominators.
Each vertex brings a factor ${\cal A}^{ac}_{\bf p,k}$, where ${\bf
p},a$ and ${\bf k},c$ are the incoming and outgoing electron momenta
and subband indices, respectively. b) Electron momenta prior ${\bf
p}$, ${\bf k}$ and after ${\bf p}+{\bf Q}$, ${\bf k}-{\bf Q}$ a
collision. The angle of incidence between momenta ${\bf p}$ and
${\bf k}$ is denoted by $\theta$, and the angle of scattering
between ${\bf p}$ and ${\bf Q}$  by $\phi$.}
\end{figure}


The general form of the optical conductivity $\sigma'(\omega,q)$ is
too cumbersome to analyze here, so we concentrate on the most
interesting, homogeneous, limit,
$\sigma'(\omega)=\sigma'(\omega,q\to 0)$. When taking the limit
$q\to 0$ it is helpful to note that the matrix (\ref{projector})
reduces to the Kronecker symbol for coinciding momenta. The matrix
element ${\cal M}_f$ thus vanishes in this limit, as required by the
presence of $q^2$ in the denominator of Eq. (\ref{dissipphon1}). We
emphasize the appearance of terms {\it linear in} $q$, which are due
to spin-orbit interaction. When the latter is absent, the
interference between the four terms in the absorption amplitude
(\ref{matrix}) leads to the cancelation of the linear terms and the
vanishing of $\sigma_2(\omega)$ \cite{Mishch}. To expand
Eq.~(\ref{matrix}) to the linear order in $q$, we note that only the
denominators need to be expanded, as the expansion of the numerators
leads only to small corrections. As a result, we obtain,
\begin{widetext}
\begin{eqnarray}
\label{conductivity} \sigma'_2(\omega)&=&e^2
\frac{\lambda^2(1-e^{-{\omega}/{T}})}{16 \omega^3} \sum_{abcd}
 \int \frac{d^2p d^2k d^2p' d^2k'}{(2\pi\hbar)^{5}}\Big|{\cal A}^{ac}_{\bf p,p'} {\cal
A}^{bd}_{\bf k,k'} V_{\bf k-k'}- {\cal A}^{ad}_{\bf p,k'} {\cal
A}^{bc}_{\bf k,p'} V_{\bf p'-k}\Big|^2
 n^a_{p}n^b_{ k}(1-n^c_{p'})(1-n^d_{k'})  \nonumber\\&&
\times \Bigl(a~{\bf n}_{\bf p}+b~{\bf n}_{\bf k}-c~{\bf n}_{\bf
p'}-d~{\bf n}_{\bf k'} \Bigr)^2 \delta (\epsilon^a_{\bf
p}+\epsilon^b_{\bf k}-\epsilon^c_{\bf p'}-\epsilon^d_{\bf k'} +
\hbar \omega) \delta({\bf p}+{\bf k}-{\bf p'}-{\bf k'} ),
\end{eqnarray}
\end{widetext}
where ${\bf n}_{\bf p}$ is the unit vector in the direction ${\bf
p}$. As seen from its form, this result is due to the interplay of
spin-orbit coupling and electron-electron interaction.

To proceed further, we utilize the fact that spin-orbit coupling is
typically small, $m\lambda \ll p_F$. Since the homogeneous optical
conductivity (\ref{conductivity}) is already proportional to
$\lambda^2$, in the leading order it is sufficient to take the limit
$\lambda \to 0$ in the delta-function and Fermi-Dirac distributions
in the integrand of Eq.~(\ref{conductivity}). The summation over the
subband indices can then be easily carried out,
\begin{eqnarray}
\label{cond_Q} \sigma'_2(\omega)= e^2
\frac{\lambda^2(1-e^{-{\omega}/{T}})}{2\omega^3}\int \frac{d^2p d^2k
d^2Q}{(2\pi)^{5}} {\cal P}({\bf p,k,Q}) \nonumber\\ \times
  n_{p}n_{ k}(1-n_{
p})(1-n_{k'}) \delta
(\epsilon_{p}+\epsilon_{k}-\epsilon_{p'}-\epsilon_{k'} + \hbar
\omega).~~
\end{eqnarray}
Here we integrated out the momentum delta-function by introducing
explicitly the momentum of electron-hole pairs, ${\bf Q}={\bf
p'}-{\bf p}={\bf k}-{\bf k'}$.  The explicit expression for the
probability of inelastic collisions ${\cal P}({\bf p,k,Q})$ is
simple but rather lengthy. In the simplest case of screened (e.g.~by
a metallic gate) short-range interaction $V_{q}=V$ this probability
is given by ${\cal P}({\bf p,k,Q})=V^2[2-({\bf n}_{\bf p}\cdot {\bf
n_k})^2-({\bf n}_{\bf p'}\cdot {\bf n}_{\bf k'})^2]$. In order to
evaluate the integrals in Eq.~(\ref{cond_Q}), it is convenient to
make use of the variables $\xi_p,\xi_k,Q,\theta,\phi$, where
$\xi_p=(p^2-p_F^2)/2m$, and the choice of angles $\theta$ and $\phi$
is illustrated by Fig.~2b). Then $d^2p d^2k d^2Q =2 (2\pi) m^2
d\xi_p d\xi_k QdQ d\theta d\phi $; here the extra factor $2$ comes
from the processes that differ from those shown in Fig.~2b) by
rotating vectors ${\bf p'}$, ${\bf k'}$  around the direction of the
vector ${\bf p}+{\bf k}$ by the angle $\pi$.

If $\omega, T \ll p^2_F/2m$, the characteristic momenta of
electron-hole pairs, $Q \sim \max(\omega,T)/v_F$, are much smaller
than the Fermi momentum $p_F$. Thus, we can approximate ${\cal
P}\approx 2 V^2 \sin^2\theta$. The argument of the delta-function in
this limit, $\omega-Qv_F\cos\phi+Qv_F\cos(\theta -\phi)$, is
independent of $\xi_p$ and $\xi_k$. This makes it possible to
perform integration over $d\xi_pd\xi_k$ first. The integral over
$dQ$ then removes the delta-function. Resulting angle integrals
cannot be calculated analytically for arbitrary temperatures.
However,  two important limits, $\omega \gg \pi T$ and $\omega \ll
\pi T$, can be easily analyzed. After some straightforward
calculations we arrive at Eq.~(\ref{result}).

{\it Long-range interaction.} Let us now address the case of a
long-range RPA Coulomb interaction which we consider here for the
$T=0$ limit only.   The interaction can be written with the help of
the usual dimensionless parameter $r_s= \sqrt{2}
me^2/(p_F\varepsilon)$, as $V_{\bf k}=\sqrt{2}\pi r_s v_F/(|{\bf
k}|+\sqrt{2}r_s p_F)$; here $\varepsilon$ is the dielectric
constant. The scattering probability can now be approximated with
\cite{clarification},
$$
{\cal P} \approx \frac{2\pi^2 r_s^2
\sin^2{\theta}}{m^2(\sqrt{1-\cos{\theta}}+r_s)^2}.
$$
The angle integrals can be performed numerically; dependence of
$\sigma'_2$ on the electronic density is shown in Fig.~3.
\begin{figure}[location=h]
\includegraphics[width=0.4\textwidth]{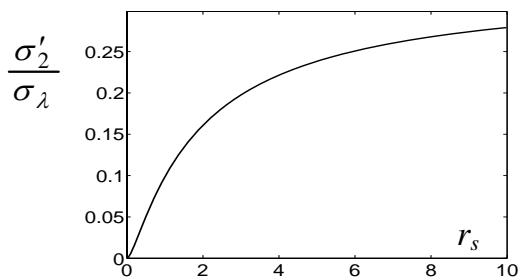}
\caption{ Dependence of the many-body optical conductivity, measured
in units of $\sigma_\lambda=e^2 \lambda^2/(2\pi v_F)^2$, on the
interaction parameter $r_s$ for long-range Coulomb interaction.}
\end{figure}

{\it Discussion}. Let us emphasize a number of important points
concerning this result. At zero temperature, the many-body
contribution $\sigma'_2$ is $\sim \lambda^2/v_F^2$ times weaker than
the one-particle Landau damping $\sigma'_1$, but it has much {\it
broader} spectrum. Also, in contrast to Landau damping $\sigma'_2$,
is enhanced with increasing the temperature. Note that when
frequency decreases, $\omega \to 0$, and temperature is kept
constant, $\sigma'_2(\omega)$ diverges. This singularity s cut-off
by the finite scattering rate $\tau^{-1}$ due to small amount of
phonons (present for any finite $T$) or impurities.

The absence of a logarithm in the many-body optical conductivity is
in a sharp contrast with other quantities describing properties of
2DEG, namely, quasiparticle lifetime \cite{Ch,HSW,JMD,ZDS}, Coulomb
drag resistivity \cite{GEM}, thermal conductivity \cite{LM,CA}, or
finite-$q$ optical conductivity \cite{Mishch}. This is a consequence
of the vanishing of the amplitude ${\cal P} \propto \theta^2$ in
Eq.~(\ref{cond_Q}) for almost collinear, $\theta \approx 0,\pi$,
scattering processes. Indeed, as the corresponding scattering
amplitudes are enhanced for such collinear processes, the
logarithmic factor, typical for 2D, may be viewed as a ``trace'' of
weakened one-dimensional singularities \cite{Chub}. The problem
analyzed in the present work, however, is inherently different. In
one dimension the discussed effect would be absent. Despite the fact
that spin-orbit coupling breaks Galilean invariance in 1D as well,
spin-conserving nature of Coulomb interaction assures that electrons
preserve their chirality (subband indices) during collisions. Thus,
the interplay of spin-orbit coupling and interactions does not
modify the optical conductivity of a one-dimensional electron
system.

An important note should be made about exchange processes. Since it
is the entire range of angles, $\theta \sim 1$, that contributes to
the optical conductivity in 2DEG with spin-orbit coupling, and not
simply the forward scattering domain, $\theta \approx 0$, the
exchange processes are important. Therefore, all diagrams in Fig.~1
are relevant. This is different from a typical scenario when
exchange processes are negligible provided that the density of
carriers is high.

{\it Summary and Conclusions}.  We have analyzed the many-body
contribution to the optical conductivity of a two-dimensional
electron liquid in the presence of spin-orbit coupling. The latter
breaks Galilean invariance, making electron dispersion
non-parabolic. This opens a possibility for optical conductivity to
be used as a probe for many-body effects. This non-trivial interplay
of spin-orbit coupling and electron-electron interactions was
revealed here for the first-time. Experimental observation of the
above effect can be performed in GaAs-based quantum wells as well as
in 2D states on the vicinal surfaces (111) of noble metals
\cite{noble,noble1}. To eradicate extraneous electron scattering the
measurements have to be performed on clean samples at low
temperatures.

We acknowledge fruitful discussions with M.~Raikh, M. Reizer,
O.~Starykh, A.~Chubukov, and S.~Gangadharaiah. The work was
supported by the Department of Energy, Office of Basic Energy
Sciences.


\begin{thebibliography}{50}

\bibitem{Pines&Noz} D. Pines and P. Nozieres, {\it The Theory of Quantum
Liquids} (Benjamin, New York, 1966).
\bibitem{BR} F.T.\ Vas'ko, JETP Lett.\ {\bf 30}, 540 (1979);
Yu.A.~Bychkov and E.I.~Rashba, J.~Phys.\ C {\bf 17}, 6039 (1984).
\bibitem{MCE} L.I. Magarill, A.V. Chaplik and M.V. \'Entin, JETP
{\bf 92}, 153 (2001).
\bibitem{MH} E.G.~Mishchenko and B.I.~Halperin, Phys.\ Rev.\ B {\bf
68}, 045317 (2003).
\bibitem{RSh} E.I.~Rashba and V.I. Sheka, in {\it Landau Level
Spectroscopy}, edited by G. Landwehr and E.I. Rashba (North-Holland,
Amsterdam, 1991), p.~131.
\bibitem{ShKhF} A. Shekhter, M. Khodas, and A.M. Finkelstein,  Phys. Rev. B {\bf 71}, 165329
(2005).
\bibitem{ChR} G.-H.\ Chen and M.E.\ Raikh, Phys.\ Rev.\ B {\bf 59},
5090 (1999).
\bibitem{Mishch} E. G. Mishchenko, M. Yu. Reizer, and L. I. Glazman, Phys.
Rev. B {\bf 69}, 195302 (2004).
\bibitem{LLVIII} L.D.~Landau and E.M.~Lifshitz, {\it Electrodynamics of
Continuous Media} (Pergamon, Oxford, 1984).
\bibitem{clarification} Note, that decreasing density (increasing
$r_s$), also means decreasing transferred momentum, $k \sim p_F$.
Thus, the condition $k<<r_s p_F$ is satisfied with progressively
better accuracy and the amplitude of scattering becomes effectively
short-range ($k$-independent).
\bibitem{Ch} A.V. Chaplik, Sov. Phys. JETP {\bf 33}, 997 (1971).
\bibitem{HSW} C. Hodges, H. Smith and J.W. Wilkins, Phys. Rev. B {\bf 4}, 302 (1971).
\bibitem{JMD} T. Jungwirth and A.H. MacDonald, Phys. Rev. B {\bf 53}, 7403 (1996).
\bibitem{ZDS} Lian Zheng and S. Das Sarma, Phys. Rev. B {\bf 53}, 9964  (1996).
\bibitem{GEM} M. Kellog, J.P. Eisenstein, L.N. Pfeiffer and
K.W. West, cond-mat/0206547.
\bibitem{LM}   A.O.\ Lyakhov and E.G.\ Mishchenko,  Phys. Rev. B {\bf
67}, 041304(R) (2003).
\bibitem{CA} G. Catelani and I.L. Aleiner, JETP {\bf 100}, 331
(2005).
\bibitem{Chub} A.V. Chubukov, D.L. Maslov, S. Gangadharaiah, and
L.I. Glazman, Phys. Rev. B {\bf 71}, 205112 (2005).
\bibitem{noble} F.J. Himpsel, {\it et al.}, Adv. Phys. {\bf 47}, 511 (1998).
\bibitem{noble1} R. N\"otzel, {\it et al.}, Nature {\bf 392}, 56
(1998); P. Segovia, {\it et al.}, {\it ibid.} {\bf 402}, 504 (1999);
P. Gambardella, {\it ibid.} {\bf 416}, 301 (2002).
\end{thebibliography}
\end{document}